\documentclass[letterpaper, 10 pt, conference]{ieeeconf}  % Comment this line out if you need a4paper

\IEEEoverridecommandlockouts                              % This command is only needed if 
\usepackage{graphicx} % for pdf, bitmapped graphics files
\usepackage{tabularx,booktabs}
\usepackage{makecell}
\usepackage{soul}
\usepackage{color}
\usepackage{amsmath} % assumes amsmath package installed

\title{\LARGE \bf
An Extreme Value Theory Approach for Understanding Queue Length Dynamics in Adaptive Corridors
}

\author{Shakib Mustavee,$^{1}$ Pushkin Kachroo (Senior Member IEEE)$^{2}$ and Shaurya Agarwal (Senior Member IEEE)$^{3}$% <-this % stops a space
\thanks{*This work was not supported by any organization}% <-this % stops a space
\thanks{$^{1}$Shakib Mustavee is a Ph.D. candidate with the Department of Civil, Environmental \& Construction Engineering, University of Central Florida, Orlando, Florida.
        {\tt\small shakib.mustavee@ucf.edu}}%
\thanks{$^{2}$Pushkin Kachroo is a Professor with the Department of Electrical \& Computer Engineering, University of Nevada, Las Vegas.
        {\tt\small pushkin@unlv.edu.}}%        
\thanks{$^{3}$Shaurya Agarwal is an Assistant Professor with the Department of Civil, Environmental \& Construction Engineering, University of Central Florida, Orlando, Florida.
        {\tt\small shaurya.agarwal@ucf.edu}}%
}

\begin{document}

\maketitle
\thispagestyle{empty}
\pagestyle{empty}

%%%%%%%%%%%%%%%%%%%%%%%%%%%%%%%%%%%%%%%%%%%%%%%%%%%%%%%%%%%%%%%%%%%%%%%%%%%%%%%%
\begin{abstract}
This paper introduces a novel approach employing extreme value theory to analyze queue lengths within a corridor controlled by adaptive controllers. We consider the maximum queue lengths of a signalized corridor consisting of nine intersections every two minutes, roughly equivalent to the cycle length. Our research shows that maximum queue lengths at all the intersections follow the extreme value distributions. To the best knowledge of the authors, this is the first attempt to characterize queue length time series using extreme value analysis. These findings are significant as they offer a mechanism to assess the extremity of queue lengths, thereby aiding in evaluating the effectiveness of the adaptive signal controllers and corridor management. Given that extreme queue lengths often precipitate spillover effects, this insight can be instrumental in preempting such scenarios. 
\end{abstract}

%\end{frontmatter}

% \linenumbers

\section{Introduction}\label{sec:intro}
Adaptive signal control technologies (ATSC) are critical for managing arterials as they can adjust signal timings according to highly variable traffic demands. ATSC adjusts the timing of red, yellow, and green lights to achieve specific operational goals, such as smooth traffic flow, maximizing throughput, ensuring access equity, and queue management. The effectiveness of adaptive strategies is determined by various measures of effectiveness (MOEs), including throughput capacity, traffic delay, and queue lengths \cite{essa2020self, shafik2017field}. Among these MOEs, queue lengths are crucial in evaluating signalized intersections' performance. Traffic signal controllers such as InSync utilize artificial intelligence algorithms to optimize traffic flow along signalized arterials. These algorithms are based on cost functions that consider the wait times for each vehicle and queue lengths for each lane. By continuously gathering real-time information from sensors and cameras, the controllers adjust the traffic light's green duration to minimize queue lengths and improve traffic efficiency. It is important to understand the evolution of queue length, particularly cycle-wise maxima, which helps identify spillover conditions, i.e., when the queue length exceeds the capacity. In queue spillover scenarios, upstream traffic conditions are affected by downstream traffic due to the mutual interactions of neighboring intersections. Therefore, estimating the cycle-based maximum queue length is crucial in optimizing signal control and managing congestion in signalized arterials. Long-term analysis of queue length provides insights into the performance and reliability of signal controllers. However, it is difficult to accurately estimate cycle-based queue length due to the lack of high-resolution data \cite{yao2019cycle}. There is a significant gap in the literature regarding the long-term behavior of cycle-to-cycle queue length. This study examines queue length dynamics in a signalized corridor and aims to uncover the long-term behavior of maximum queue length by probabilistic analysis. We draw the maximum queue length of each cycle and characterize its distribution in the light of the extreme value theory (EVT), which is a statistical theory that deals with extreme values and events that occur at the tails of probability distributions. According to EVT, the largest value drawn from a sample of a given size can follow one of three possible distributions: Weibull, Fréchet, or Gumbel distributions. Since uncommon and extreme conditions can cause significantly greater consequences, understanding and quantifying rare occurrences is essential for reliability analysis and impact assessment. EVT aims to assess the likelihood of events surpassing previous extremes based on ordered samples of random variables, and it focuses on studying extreme deviations from distribution medians. Extreme value analysis is widely applicable across fields such as structural engineering, finance, economics, geology, telecommunication, and traffic safety. It is crucial for comprehending and preparing for extreme events across various domains. In transportation, the primary application of EVT remains in surrogate safety measures based on traffic safety analysis. Studying the queue length dynamics in the light of extreme value theory is intriguing as it provides insight into extreme incidents when queue lengths exceed capacity.

To our best knowledge, this is the first research paper that uses EVT to analyze queue length dynamics. The contributions of the paper are as follows:

\begin{itemize}
    \item The paper introduces extreme value analysis of queue length time series.
    \item It provides mathematical justifications for the emergence of extreme value distribution from queueing theory. 
    \item It provides empirical evidence for the emergence of extreme value distribution using real data.  
    %\item EVT characterization of queue lengths can potentially help evaluate the performance of an adaptive signal control. % This is NOT a contribution.
\end{itemize}

The structure of the paper is as follows. Section~\ref{sec:litrev} discusses the relevant literature on EVT analysis in transportation engineering and general queueing systems. Section~\ref{sec:mathback} provides background on mathematical tools, while Section~\ref{sec:res} provides a rich discussion on data and obtained results.

%\textbf{Hypothesis:} A cycle's queue length is defined by that cycle's maximum queue length. Therefore, the maximum queue length distribution is expected to follow some extreme value distribution- (i) Gumbel, (ii) Frachet (iii) Weibull

%this is taken from (1) https://en.wikipedia.org/wiki/Extreme_value_theory#:~:text=Extreme%20value%20theory%20is%20used,as%20the%201755%20Lisbon%20earthquake. (2) https://schumacher.atmos.colostate.edu/gherman/EVT_V1P1.pdf

\section{Literature Review} \label{sec:litrev}
EVT has been used in traffic safety research for over two decades, particularly to analyze the association between traffic conflicts and crashes. A thorough analysis of the studies that have applied EVT in traffic safety can be found in \cite{ali2023assessing}. The review paper offers elaborate discussions on assessing alternate modeling methodologies and associated issues. There are only a few examples of EVT being applied in transportation literature, apart from safety applications. One example is the use of an EVT-based approach to measure extremely prolonged travel time and assess potentially influential factors of emergency vehicles, as proposed in \cite{zhang2019analyzing}. Another example is the use of EVT in traffic incident detection, as described in \cite{tuli2022radnet}. However, to the best of our knowledge, EVT-based approaches have not been used in queue length analysis. While transportation literature largely ignores extreme value analysis of queue length analysis, extensive research exists regarding the extreme value distributions of queue lengths and waiting time in various other queuing systems related to operations research and telecommunication \cite{asmussen1998extreme, berger1995maximum, akbash2024extreme}. These works primarily focus on deriving explicit equations and limiting cases for various queuing conditions.

%https://www.sciencedirect.com/science/article/pii/S0968090X18303449?casa_token=Xq4m6P1CEW0AAAAA:S3VpFPDO_zwEp-Ds9RuobLsZ4VOyqhyd5mwpLriyZxwUSQjGEUzmcthYQDtWzHUo8RBjTrFLgms

%https://ieeexplore.ieee.org/stamp/stamp.jsp?tp=&arnumber=9714796

%https://ieeexplore.ieee.org/stamp/stamp.jsp?tp=&arnumber=1378217

% Cycle level conflict: Cycle-level traffic conflict prediction at signalized intersections with LiDAR data and Bayesian deep learning

% Real-Time multi-objective optimization of safetyand mobility at signalized intersections

%Incident prediction: https://arxiv.org/pdf/2206.05602

\section{Mathematical Background} \label{sec:mathback}

\subsection{Generalized Extreme Value Distribution}
%%%%%%%%%%%%%%%%%%%%%%%%%%%%%%%%%
EVT focuses on modeling the maximum of a sample of identical and independent random variables (i.i.d r.v). The theory says that after proper renormalization, their distribution can only converge to one of only 3 possible distribution families: type-I (Gumbel), type-II (Fréchet), and type-III (Weibull). The Fisher-Tippett-Gnedenko theorem (extreme value theorem) provides the criteria for the maximum value of i.i.d r.v) to converge to a Generalized Extreme Value (GEV) distribution characterized by three parameters: \( \mu, \sigma, \) and \( \xi \) which combines the Gumbel, Fréchet, and Weibull families under one general distribution. It means that under different conditions, these distributions emerge from the GEV distribution.  

We consider the set of i.i.d r.v: \[ X_1, X_2, \ldots, X_n \]

Here, we sample each \( X_i \) from any function \( F \). The distribution of the function is not known. Let \( \Gamma_n = \max(X_1, X_2, \ldots, X_n) \).

Let there exists a set of real numbers \( \alpha_n, \beta_n \), where \( \alpha_n \geq 0 \), such that \( \frac{\Gamma_n - \beta_n}{\alpha_n} \) has a limiting distribution in the limit of large \( n \) and it is non-degenerate. Then extreme value theorem says:

\[ \Gamma_n \sim GEV(\mu, \sigma, \xi) \]

This means the maximum (and minimum) of the i.i.d r.v follows generalized extreme value distribution (GEV).

\textbf{Generalized Extreme Value (GEV) Distribution:}
First, we discuss the cumulative distribution function (CDF) of the Generalized Extreme Value (GEV) distribution and then we see how it is related to the three different types of extreme value distributions. The CDF of GEV is:
\[ F(x;\mu,\sigma,\xi) = \exp\left\{-\left[1 + \xi\left(\frac{x-\mu}{\sigma}\right)\right]^{-1/\xi}\right\}, \]
where, \( \mu \) is the location parameter, \( \sigma \) is the scale parameter, \( \xi \) is the shape parameter.

\textbf{Type-I or Gumbel Extreme Value Distribution:}
The Gumbel distribution emerges when \( \xi = 0 \) in the GEV distribution:
\[ F_{\text{Gumbel}}(x;\mu,\sigma) = \exp\left\{-\exp\left[-\left(\frac{x-\mu}{\sigma}\right)\right]\right\}, \]
where \( \mu \) and \( \sigma \) are the location and scale parameters, respectively.

\textbf{Type-II or Fréchet Extreme Value  Distribution:}
The Fréchet distribution emerges when \( \xi > 0 \) in the GEV distribution:
\[ F_{\text{Fréchet}}(x;\mu,\sigma,\xi) = \exp\left\{-\left[1 + \xi\left(\frac{x-\mu}{\sigma}\right)\right]^{-1/\xi}\right\}, \]
where \( \mu \), \( \sigma \), and \( \xi \) are the location, scale, and shape parameters, respectively.

\textbf{Type-III or Extreme Value (Weibull) Distribution:}
The Weibull distribution emerges when \( \xi < 0 \) in the GEV distribution:
\[ F_{\text{Weibull}}(x;\mu,\sigma,\xi) = \exp\left\{-\left[1 + \xi\left(\frac{x-\mu}{\sigma}\right)\right]^{-1/\xi}\right\}, \]
where \( \mu \), \( \sigma \), and \( \xi \) are the location, scale, and shape parameters, respectively.

% The following section is written with the help of chatGPT

There are two methods for modeling extreme values within a random process. The first approach, the block minima/maxima, involves dividing the data into non-overlapping, continuous, and uniform intervals. Within each interval, either the maximum or the minimum entry is recorded. These minima/maxima within the intervals follows a generalized extreme value (GEV) distribution, as described by Gumbel in 1958 \cite{gumbel1958statistics}. The second approach, referred to as threshold exceedances or the peak over the threshold approach, involves selecting observations surpassing/falling short of a threshold value. Using this method, the observations chosen are typically approximated using a generalized Pareto distribution (GPD), as outlined by Pickands III in 1975 \cite{pickands1975statistical}.

While the block maxima approach is generally easier to apply and interpret, it may not always be directly applicable and often leads to inefficient use of data. Conversely, threshold exceedance can potentially lead to more efficient data utilization, but it is more challenging to implement and ensure that the underlying conditions of the theory are met.

%There are two approaches to identifying and modeling the extreme values of a random process: (i) the block Maxima approach, which divides the observation period into nonoverlapping, continuous, and equal intervals and collects the maximum entries of each interval (Gumbel, 1958). Maxima from these blocks (intervals) can be fitted into a generalized extreme value (GEV) distribution.(ii) threshold exceedances, i.e., the peak over the threshold approach selects the observations that exceed a certain high threshold. A generalized Pareto distribution (GPD) is usually used to approximate the observations selected with the POT approach. (Pickands III, 1975). In general, the block Maxima approach is Simple to apply and interpret. However, the Framework might not as often directly useful and might not make the most efficient use of time series. On the contrary, threshold exceedance can be the efficient use of data, but it is harder to implement and to know that the conditions of the theory have been satisfied.

% \Study materials- https://www.simtrade.fr/blog_simtrade/extreme-value-theory-block-maxima-peak-over-threshold/

% https://georgebv.github.io/pyextremes/user-guide/2-extreme-value-types/

% https://en.wikipedia.org/wiki/Seasonal_adjustment

% https://schumacher.atmos.colostate.edu/gherman/EVT_V1P1.pdf

%%%%%%%%%%%%%%%%%%%%%%%%%%%%%%%%
\subsection{Queuing Process of a Signalized intersection}
For several decades, signalized intersections have been modeled as a queueing system. Queueing models can estimate the average queue length and waiting time in a signalized intersection. Different models, such as M/M/1, M/G/1, GI/M/1, and M/D/1 have been proposed to model various attributes of signalized intersections in different scenarios. For example, Pollaczek-Khintchine equation is widely used in transportation literature to model average queue length and delay assuming intersections as an M/G/1 system \cite{rouphail151992traffic}. A review of various queueing models used in modeling intersections is provided in \cite{gunes2021smart}. Traditional queueing models can not fully capture the features of a signalized intersection due to the alteration of signals. During the red time, signalized intersections suspend service. The queuing process of vehicles at signals can be characterized by queueing models with server vacations \cite{yang2018evolution}. Despite their limitations, these models provide a workable approximation of real queue lengths and delays at signalized intersections. %Queue formation of signalized intersection
 
%https://link.springer.com/chapter/10.1007/978-3-319-61428-1_10

%%%%%%%%%%%%%%%%%%%%%%%%%%%%%%%%%

%The sequence \( \{X_t\}_{t \geq 0} \) is considered regenerative if there exists a renewal process, possibly with some delay, characterized by epochs \( 0 < T_0 < T_1 < T_2 < \ldots \), such that the cycles \( \{X_t \mid t \in [T_i, T_{i+1}) \} \), for \( i \geq 0 \), are independent and share the same distribution for \( i \geq 1 \) (with \( T_{-1} \) conventionally set to 0) \cite{asmussen1998extreme}. It's worth noting that the initial cycle, corresponding to \( i = 0 \), might have a distinct distribution, as is the case in a stationary regenerative process. The standard cycle corresponds to \( i = 1 \), and we denote its length as \( t = T_1 - T_0 \). We define \( m = E(t) \), the expected value of \( t \), and \( X_i(t) = \max_{T_{i-1} \leq s \leq T_i} X_s \), where \( X_t \) represents the maximum value within the generic cycle \( t \). Then \( X_1(t), X_2(t), \ldots \) are independent and identically distributed (i.i.d.) and share the distribution of \( X_t \). Since the number of cycles completed by time \( t \) asymptotically approaches \( t/m \), we anticipate that \( X_t \) is reasonably close to \( \max_{1 \leq i \leq t/m} X_i(t) \), the asymptotic behavior of which can be inferred from i.i.d. theory (assuming the tail behavior of \( X_t \) is known).

%%%%%%%%%%%%%%%%%%%%%%%%%%%%%%%%
We use queueing theory to explain the mathematical structure of cycle maxima. Let's assume a signalized intersection with a stable queueing formation and appropriate initial distribution. Here, \( W_i \) represents the waiting time of the \( i^{th} \) vehicle, and \( L(t) \) denotes the queue length at time \( t \). Here, the objective is to identify approximate distributions for the maximum queue lengths and waiting associated with this process. For sufficiently large values of \( i \) and \( t \) we assume

\[ W_i^* = \max \{W_k : 0 \leq k \leq i\}, \quad i \geq 0, \]

and

\[ L^*(t) = \max \{L(s) : 0 \leq s \leq t\}, \quad t \geq 0, \]

Now, according to extreme-value limit theorems, the maximum queue length process \( L^*(t) \), that the approximations should follow the mathematical structure \cite{berger1995maximum}:

\[ L^*(t) \sim \gamma(\log t + \log \beta + Z), \]

where \( t \) is presumed to be relatively large, \( \log \) denotes the natural logarithm, \( Z \) follows the Gumbel cumulative distribution function (type-I extreme-value CDF):

\[ P(Z \leq x) \equiv \Lambda(x) = \exp (-e^{-x}), \quad -\infty < x < \infty, \]

and \( \gamma \) and \( \beta \) are positive constants that depend on the particular process being analyzed.

%%%%%%%%%%%%%%%%%%%%%%%%%%%%%%%%%%%%%%%%%%%%%%%%%%%%%%%%%%%%%%%%%%%%%%%%%%%
\section{Case Study} \label{sec:res}
\subsection{Data Description}
\begin{figure}[!htbp]\center
\includegraphics[width=1.0\linewidth]{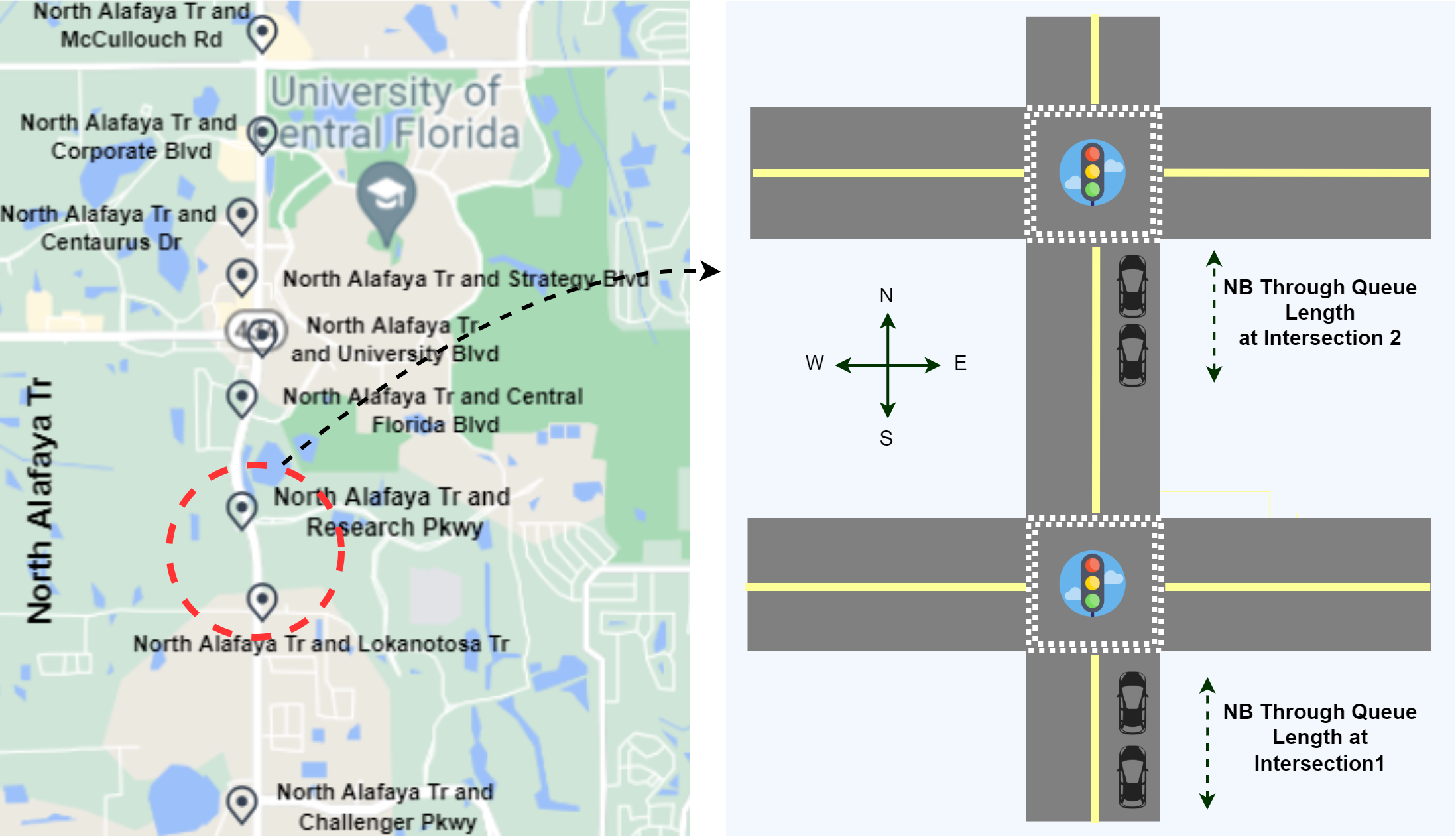}
%\vspace{-4mm}
\caption{Location of intersections on Alafaya corridor (left) and illustration of the corridor as a dynamical system (right). Northbound (NB) through queue length at each intersection is measured in distance units.}
\vspace{-4mm}
\label{fig:corridor}
\end{figure}

The dataset comprises queue lengths collected from \textit{InSync} adaptive traffic signals recorded between December 18, 2017, and February 14, 2018, at the Alafaya Trail (SR-434) corridor in East Orlando, FL, shown in Figure~\ref{fig:corridor}. This study focuses on queue lengths at nine signalized intersections along this corridor, primarily investigating northward movements (NT) since it is the major road. Prior research demonstrated the effectiveness of adaptive signals for major roads \cite{shafik2017field}. Traffic sensors may not always produce accurate data due to various factors, such as detector malfunctions, errors during data storage, and unfavorable weather conditions. The study followed the outlier mitigation and data noise reduction methods suggested in \cite{rahman2021real} and \cite{das2023koopman}. 

\subsection{Data Processing}
The extreme value theory assumes i.i.d r.v (independent and identically distributed random variables), meaning that the random variables are- (i) {independent:} the samples are independent. (ii) {identically distributed:} each sample comes from the same distribution. EVT often assumes stationarity of the underlying stochastic process. However, the traffic flow dynamics exhibit periodic behavior imposed by peak hours, circadian and weekly cycles. For example, the hybrid model proposed in \cite{zhang2018hybrid} for traffic flow prediction decomposes time series into four different parts- a periodic component, a trend component, a stationary component, and a volatility component. In \cite{das2023koopman}, the queue length time series was decomposed into quasiperiodic and chaotic components. Therefore, seasonality from the traffic flow time series must be adjusted before applying EVT.

% Seasonality removing 
%Assumes i.i.dRVs (Independent and identically distributed random variables) Independent: Each sample is independent from all other samples (big assumption!) Identically Distributed: Each sample is taken from the same underlying distribution. Assumes no seasonality in the time series (big assumption!) Can correct for seasonality; not always a major concern Assumes stationarity in the time series (big assumption!) Can correct for known trends Random: Value sampled from a probability distribution, varies due to chance: Measurements are unbiased estimators of the desired quantity

% https://machinelearningmastery.com/time-series-seasonality-with-python/#:~:text=A%20simple%20way%20to%20correct,the%20value%20from%20last%20week.

% https://en.wikipedia.org/wiki/Seasonal_adjustment

% Irregular variations: https://link.springer.com/referenceworkentry/10.1007/978-0-387-32833-1_204
% https://digfir-published.macmillanusa.com/psbe4e/psbe4e_ch13_2.html
% https://web.vu.lt/mif/a.buteikis/wp-content/uploads/2019/02/Lecture_03.pdf
% Try this code: https://medium.com/@venujkvenk/exploring-time-series-data-unveiling-trends-seasonality-and-residuals-5cace823aff1
% https://otexts.com/fpp2/residuals.html
% Seasonal decomposition: https://timeseriesreasoning.com/contents/time-series-decomposition/

%%%%%%%%%%%%%%%%%%%%%%%%%%%%%%%%%%%%%%%%%%%%%%%%%
In signal processing, a time series is often decomposed as: 
\[ X_{t} = S_{t} + T_{t} + C_{t} + R_{t} \]
where \( X_{t} \) represents the data at time \( t \),

Here, four key components of time series comprise are:

\noindent - ($S_{t}$) signifies the periodicity or the seasonal component  \\
\noindent-  ($T_{t}$) is the gradual changes or the trend component \\
\noindent- ($C_{t}$) denotes the cyclical component  \\
\noindent- ($E_{t}$) indicates the residual part of the time series

\textbf{Distinguishing seasonal and cyclic patterns:}
Seasonal patterns have a fixed length, while cyclic patterns are variable. Cyclic patterns occur when data fluctuates unpredictably. The average cycle is longer than seasonality, and the magnitude fluctuations are more variable.

% taken from: https://en.wikipedia.org/wiki/Seasonal_adjustment

\begin{figure}
        \includegraphics[width=0.95\linewidth]{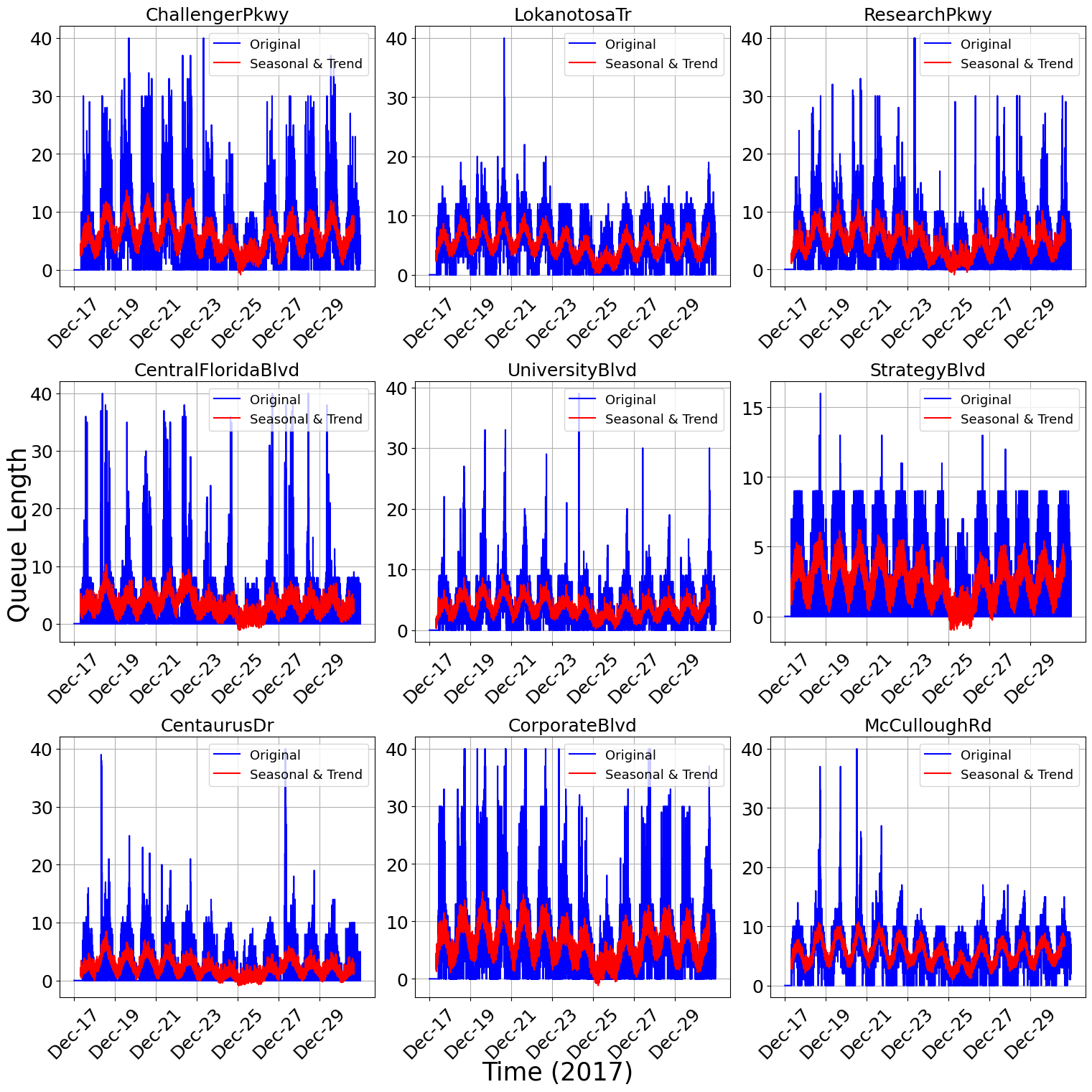} % Replace 'image2.png' with your image file name
        \caption{Illustrates the original queue length time series and adjusted time series after removing seasonal and trend components from the original queue length time series.}
        \label{fig:timeseries_adj_original}
    \end{figure}

\begin{figure}
        \includegraphics[width=0.95\linewidth]{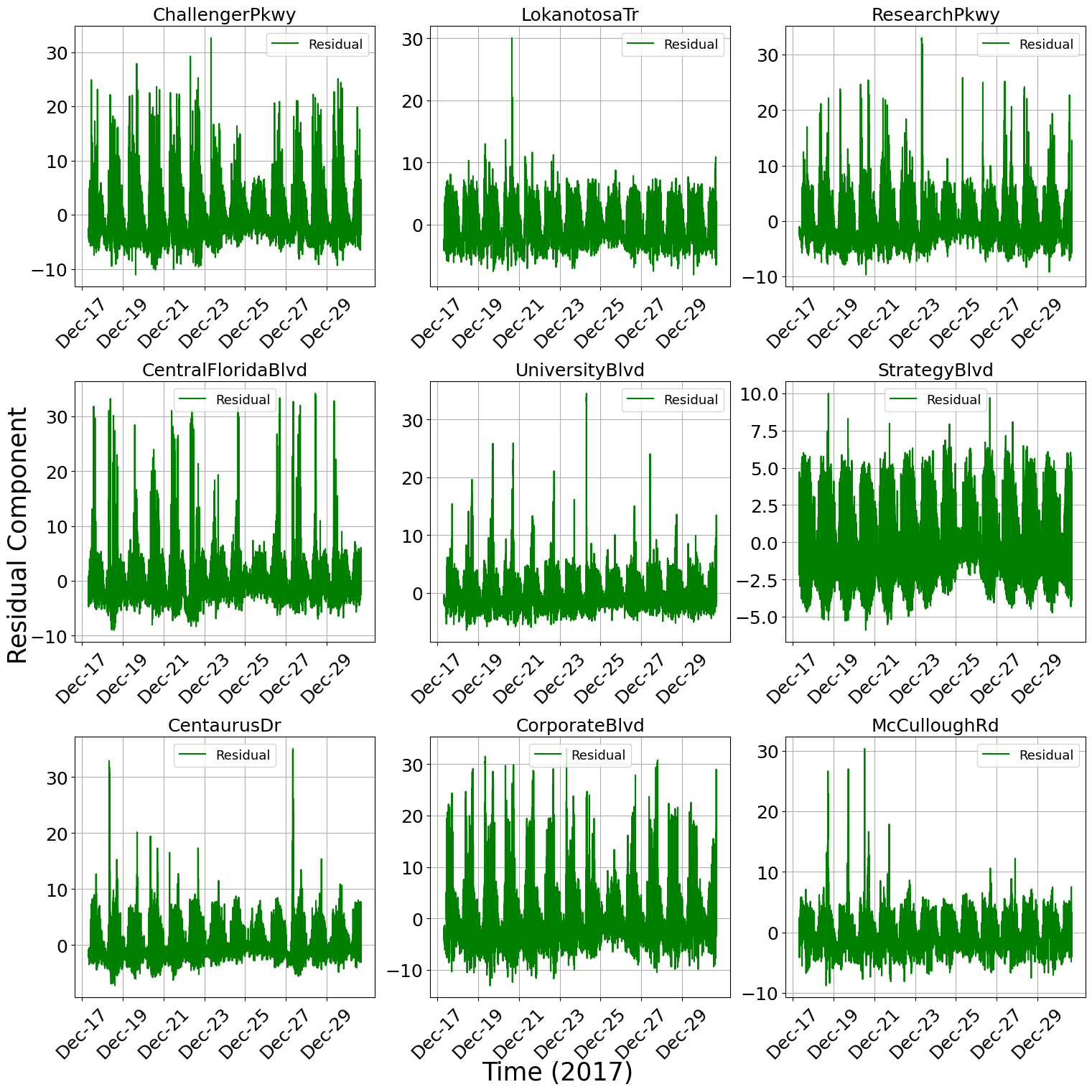} % Replace 'image2.png' with your image file name
        \caption{Illustrates the residual part computed after removing seasonal and trend components from the original queue length time series.}
        \label{fig:residue}
    \end{figure}

We observe the distribution of residual components of the queue length time series. Together, seasonality and trend are often known as autoregressive behavior. To obtain residual components, we remove autoregressive behavior from the time series using seasonal\textunderscore decompose, a Python-based software \cite{sesonal}. It is important to note that the queue length time series dynamics have multiple driving frequencies \cite{das2023koopman}. The periodic component of the queue length time series has several frequencies. Therefore, removing the seasonality of queue length is not straightforward. Traffic flow dynamics are driven by a circadian circle. Hence, it is a common practice to assume the seasonality period of traffic flow as 24 hours. However, during late night and early morning off-peak hours (10 PM to 07 AM), the queue length usually remains zero. Therefore, we assume the period of seasonality in queue length time series as 15 hours. Figure~\ref{fig:timeseries_adj_original} illustrates the original queue length time series (shown in blue) along with the seasonal and trend components. The seasonal and trend components are added together (shown in red). Figure~\ref{fig:residue} shows the residual part of the time series. 

\subsection{Distribution of Residual Component}

It is observed that the residual components do not follow normal distribution. We tried to find the best-fit distribution for the residual part and used \textit{distfit}, a Python package, to find the best-fit distribution \cite{Taskesen_distfit_is_a_2020}. The package can test a series of distributions fitted on the empirical data and calculate the residual sum of squares (RSS) for each distribution. RSS quantifies the variance between predicted and observed values in empirical data. Essentially, it highlights the disparities in estimations. It serves as a metric of how well a model aligns with the actual data. A lower RSS suggests a closer model alignment with the data points. The RSS calculation is represented by the formula:

\[ RSS = \sum_{i=1}^{n}(y_i - f(x_i))^2 \]

Here, \( y_i \) denotes the $i^{th}$ value of the target variable, \( x_i \) is the $i^{th}$ value of the data, and \( f(x_i) \) denotes the predicted value of \( y_{i} \). Here, the best-fitting distribution is determined based on the lowest RSS score.       

\begin{table*}[!htbp]

\centering
\resizebox{\textwidth}{!}{%
\begin{tabular}{ccccccccccc}
\hline
%\multicolumn{1}{c}{} & \multicolumn{1}{c}{} & \multicolumn{6}{c}{\textbf{Intersection Name}}\\

\cline{1-9}
\hline
Rank & \textbf{Challenger Pkwy}& \textbf{Lokanotosa Tr}& \textbf{Research Pkwy} & \textbf{Central FL Blvd} & \textbf{University Blvd} & \textbf{Strategy Blvd} & \textbf{Centaurus Dr} & \textbf{Corporate Blvd} & \textbf{McCullough Rd}\\
\hline 
1 & \makecell{genextreme \\ (0.002835)} & \makecell{vonmises\_line \\ (0.00782)} & \makecell{Fatigue life \\ (0.003468)} & \makecell{johnsonsu \\ (0.003118)} & \makecell{mielke \\ (0.00516)} & \makecell{beta \\ (0.03804)} & \makecell{johnsonsu \\ (0.003733)} & \makecell{johnsonsu \\ (0.001046)} & \makecell{lognorm \\ (0.010334)} \\  
\hline
2 & \makecell{invweibull \\ (0.002835)} & \makecell{truncnorm, norm \\ (0.008892)} & \makecell{recipinvgauss \\ (0.003489)} & \makecell{burr \\ (0.003422)} & \makecell{alpha \\ (0.005407)} & \makecell{chi, nakagami \\ (0.038059)} & \makecell{exponnorm \\ (0.005357)} & \makecell{exponnorm \\ (0.004448)} & \makecell{genlogistic \\ (0.010349)} \\        
\hline
3 & \makecell{powerlognorm \\ (0.002977)} & \makecell{loggamma \\ (0.009469)} & \makecell{lognorm \\ (0.003503)} & \makecell{mielke \\ (0.003619)} & \makecell{fisk \\ (0.006609)} & \makecell{gengamma \\ (0.038083)} & \makecell{mielke \\ (0.006891)} & \makecell{mielke \\ (0.004875)} & \makecell{gumbel\_r \\ (0.010366)} \\
\hline
4 & \makecell{invgamma \\ (0.002985)} & \makecell{rdist \\ (0.009515)} & \makecell{johnsonsb \\ (0.003512)} & \makecell{exponnorm \\ (0.005025)} & \makecell{genextreme \\ (0.00697)} & \makecell{pearson3, chi2,\\ gamma \\ (0.038166)} & \makecell{burr \\ (0.007078)} & \makecell{t \\ (0.009079)} & \makecell{betaprime \\ (0.010409)} \\   
\hline
5 & \makecell{weibull\_max \\ (0.003162)} & \makecell{nakagami \\ (0.009756)} & \makecell{betaprime \\ (0.003621)} & \makecell{genextreme \\ (0.010213)} & \makecell{exponnorm \\ (0.007005)} & \makecell{betaprime \\ (0.038412)} & \makecell{alpha \\ (0.01051)} & \makecell{fisk \\ (0.009129)} & \makecell{johnsonsb \\ (0.010677)} \\   
\hline
6 & \makecell{gumbel\_r \\ (0.003188)} & \makecell{chi \\ (0.009756)} & \makecell{f \\ (0.003626)} & \makecell{invweibull \\ (0.010213)} & \makecell{genlogistic \\ (0.007018)} & \makecell{f \\ (0.03888)} & \makecell{genextreme \\ (0.010805)} & \makecell{loglaplace \\ (0.00917)} & \makecell{recipinvgauss \\ (0.010912)} \\      \hline   
7 & \makecell{genlogistic \\ (0.003227)} & \makecell{foldnorm \\ (0.009906)} & \makecell{invgamma \\ (0.003634)} & \makecell{genlogistic \\ (0.010808)} & \makecell{gumbel\_r \\ (0.007145)} & \makecell{recipinvgauss, \\fatiguelife \\ (0.038965)} & \makecell{invweibull \\ (0.010805)} & \makecell{tukeylambda \\ (0.009183)} & \makecell{fatiguelife \\ (0.010938)} \\
\hline
8 & \makecell{johnsonsu \\ (0.003352)} & \makecell{logistic \\ (0.010683)} & \makecell{genextreme \\ (0.003854)} & \makecell{gumbel\_r \\ (0.011343)} & \makecell{weibull\_max \\ (0.007481)} & \makecell{invgauss \\ (0.039039)} & \makecell{exponweib \\ (0.011539)} & \makecell{alpha \\ (0.01008)} & \makecell{invgauss \\ (0.010942)} \\   
 \hline
9 & \makecell{lognorm \\ (0.003538)} & \makecell{t \\ (0.01083)} & \makecell{invweibull \\ (0.003854)} & \makecell{betaprime \\ (0.013841)} & \makecell{burr \\ (0.007718)} & \makecell{powerlognorm \\ (0.039338)} & \makecell{genlogistic \\ (0.011597)} & \makecell{invweibull \\ (0.010131)} & \makecell{invgamma \\ (0.011627)} \\  
\hline
10 & \makecell{johnsonsb \\ (0.003822)} & \makecell{dweibull \\ (0.010993)} & \makecell{gumbel\_r \\ (0.003927)} & \makecell{invgamma \\ (0.013917)} & \makecell{invgamma \\ (0.008051)} & \makecell{gumbel\_r, \\weibull\_max, \\ (0.039361)} & \makecell{gumbel\_r \\ (0.011838)} & \makecell{genextreme \\ (0.010131)} & \makecell{powerlognorm \\ (0.011706)} \\
% Continue adding cells in the same format...
\hline
\end{tabular}%
}
\caption{Ranking of Best-fit Distributions Based on RSS Error}
\label{table:1}
\end{table*}

In the analysis, we tested a total of 79 different distributions for each intersection, and Table~\ref{table:1} summarizes the results. The table shows the ranking of the best-fitted distributions based on their RSS scores. Each column lists the best-fit distributions for each intersection, with the corresponding RSS score shown in parentheses. Note that we used acronyms for distribution names in the table. The full form can be found in \cite{Taskesen_distfit_is_a_2020}. 

We observed that except for Lokanotosa Tr, the distributions of the rest of the intersections could be described by any of the extreme value distributions such as genextreme (generalized extreme value distribution), gumbel\textunderscore r (Gumbel right), gumbel\textunderscore l (Gumbel left), weibull\textunderscore max (Weibull maximum), invweibull (inverse Weibull). This suggests that the distribution of cycle-wise maximum queue length of the corridor follows extreme value distributions. 

\begin{figure}
       \includegraphics[width=0.95\linewidth]{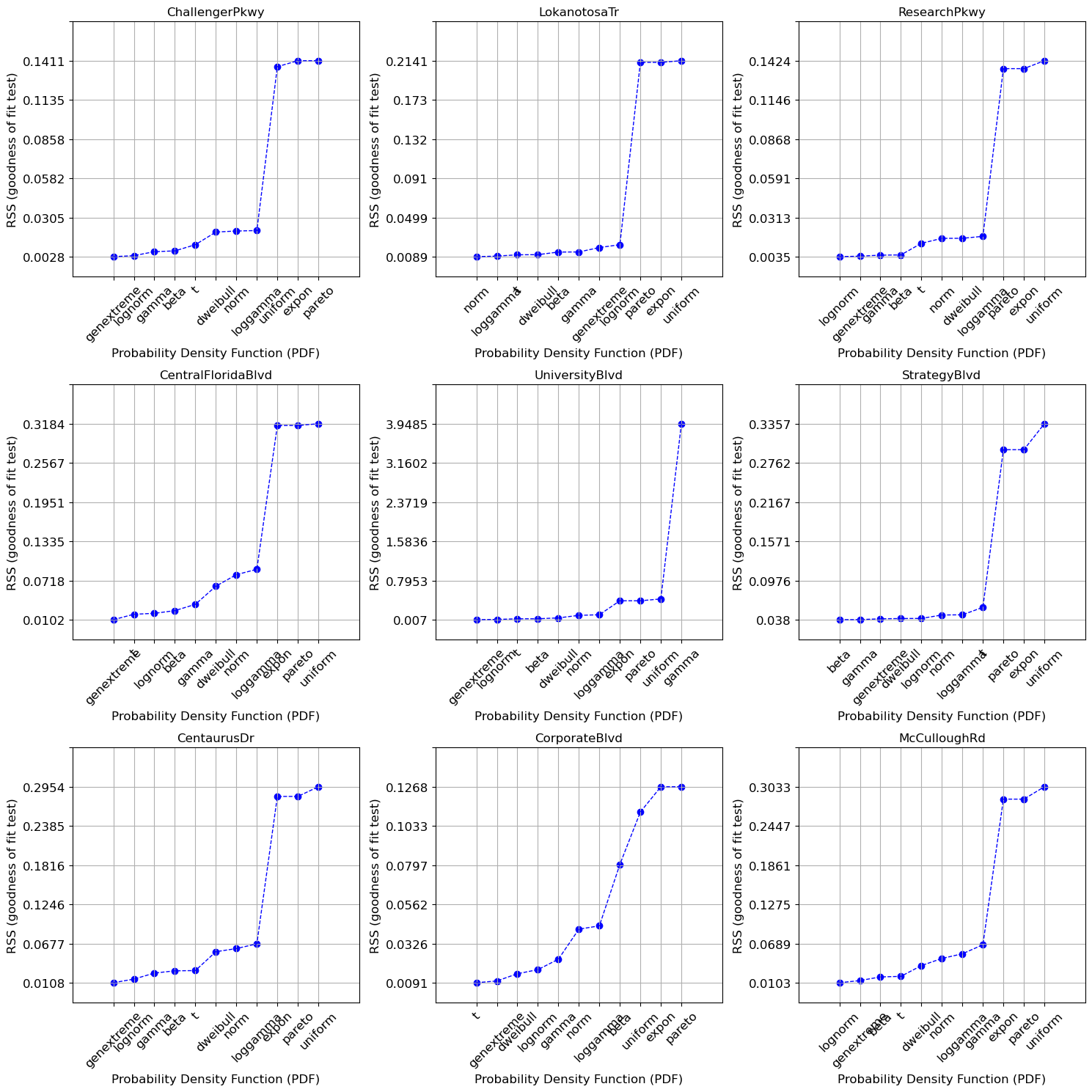} % Replace 'image2.png' with your image file name
        \caption{Illustrates the RSS score of some popular distributions. The general extreme value distribution is the best fit for most intersections.}
        \label{fig:comparison_plot}
    \end{figure}

Figure~\ref{fig:comparison_plot} presents a visual representation of the RSS score of various commonly used distributions, including generalized extreme value, lognormal, t, loggamma, beta, exponential, Pareto, normal, double Weibull, and uniform. Upon examining the figure, we observe that the generalized extreme value distribution has the best RSS score in five intersections and the second-best RSS score in two intersections when compared to the other popular distributions. In the remaining intersections, the RSS score of the generalized extreme value distribution is comparable to the fitted distributions with the best scores.   

\begin{figure}
       \includegraphics[width=0.95\linewidth]{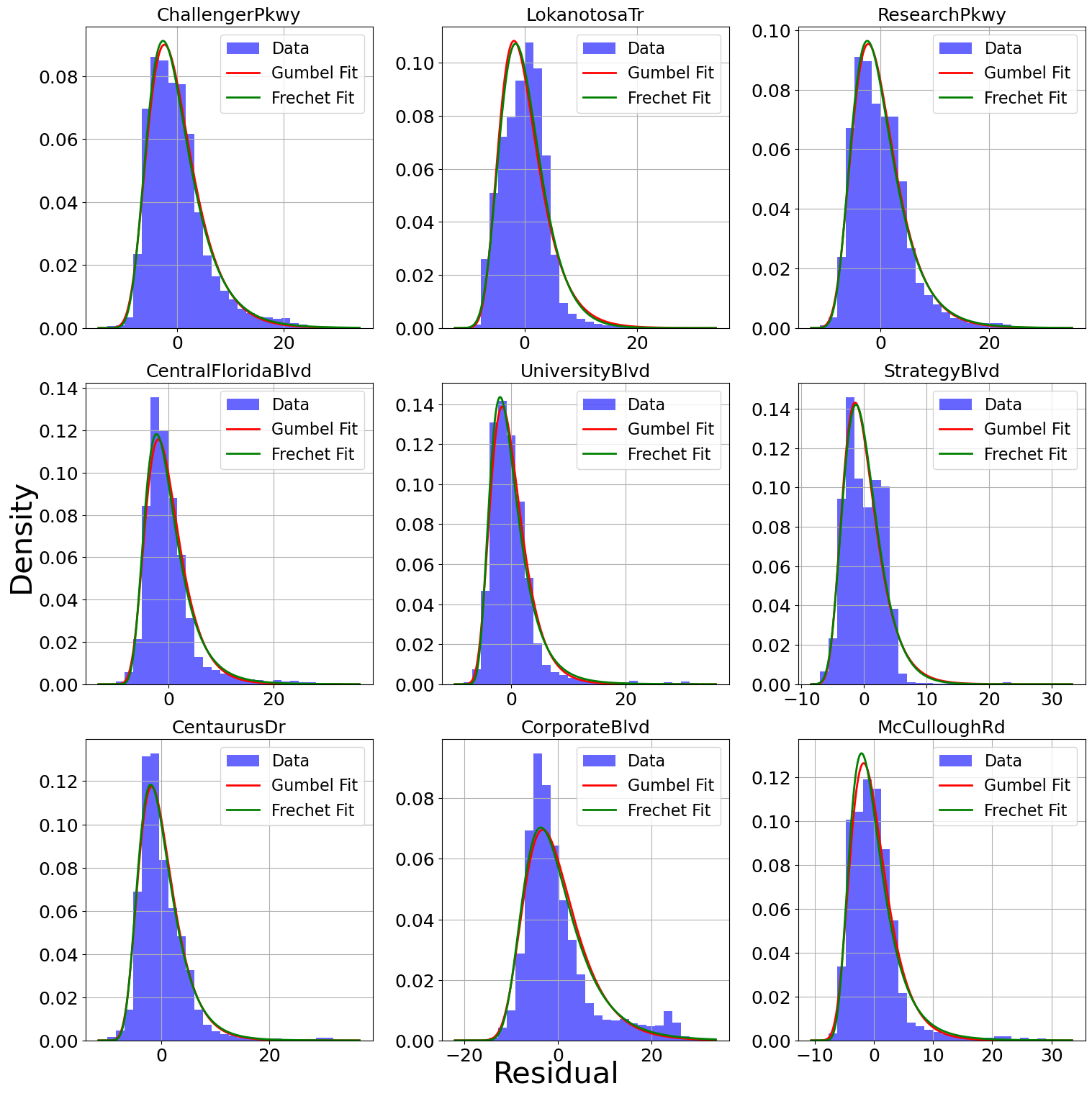} % Replace 'image2.png' with your image file name
        \caption{Illustrates the probability distribution of the residual part of the queue length time series. Gumbel and Fréchet distributions are fitted into the histogram}
        \label{fig:distribution_plot}
    \end{figure}

The histogram of the residual is displayed in Figure~\ref{fig:distribution_plot}. We fitted two distributions to the histogram: Gumbel and Fréchet, which are special cases of the generalized extreme value distribution. Visually, both distributions fit the histogram well. However, to further verify this, we applied a few graphical tests.   

The Probability-Probability (P-P) plots and Quantile-Quantile (Q-Q) plots are two graphical tools used in data analysis to assess the similarity between two distributions. The P-P plot is utilized to evaluate quality of match between a given data set and a particular probability distribution. It compares the empirical CDF of the data with the hypothesized actual CDF. If the scatter plot of the two distributions exhibit a roughly linear pattern, it suggests that the hypothesized true distribution is a good fit for the data set being analyzed. On the contrary, the Q-Q plot is a graph used to identify the true probability distribution of a given set of data in a similar way to P-P plots. However, this graph compares the quantiles of the empirical distribution of the data with the quantiles of the assumed true probability density function. If the resulting graph of these two distributions follows a linear pattern, it suggests that the assumed probability density function reasonably matches the given data.

Figure~\ref{fig:pp} and Figure~\ref{fig:qq} {present the P-P and Q-Q plots for all intersections respectively. These plots assume that the residual follows the Gumbel distribution. As per the figures, it is evident that all intersections follow a linear pattern in their respective P-P and Q-Q plots. This indicates that using the Gumbel distribution is a suitable way to approximate the queue length time series of the corridor.  
    \begin{figure}
        \includegraphics[width=0.95\linewidth]{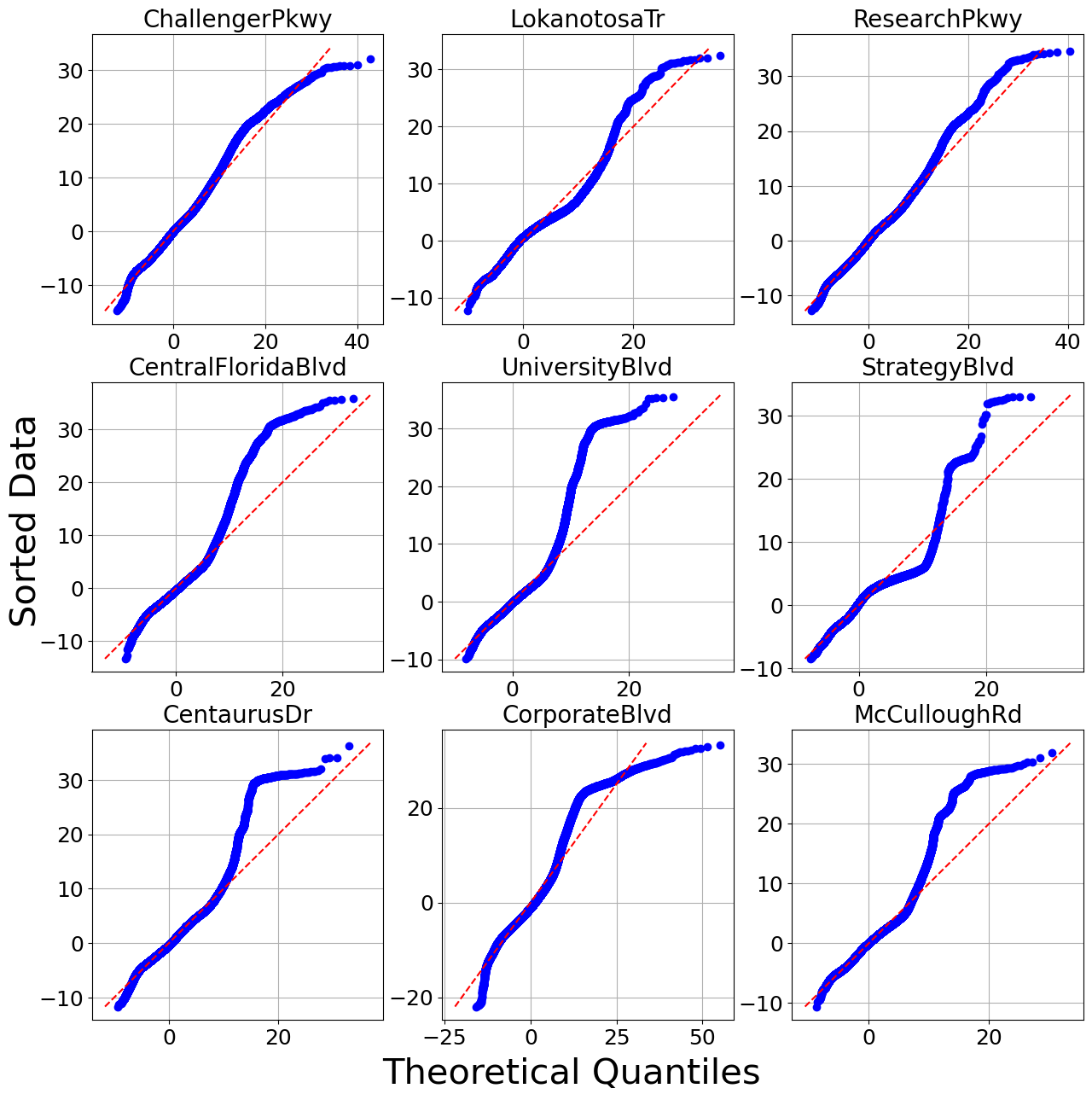} % Replace 'image2.png' with your image file name
        \caption{P-P plots of the data assuming it follows Gumbel distribution}
        \label{fig:pp}
    \end{figure}

     \begin{figure}
        \includegraphics[width=0.95\linewidth]
        {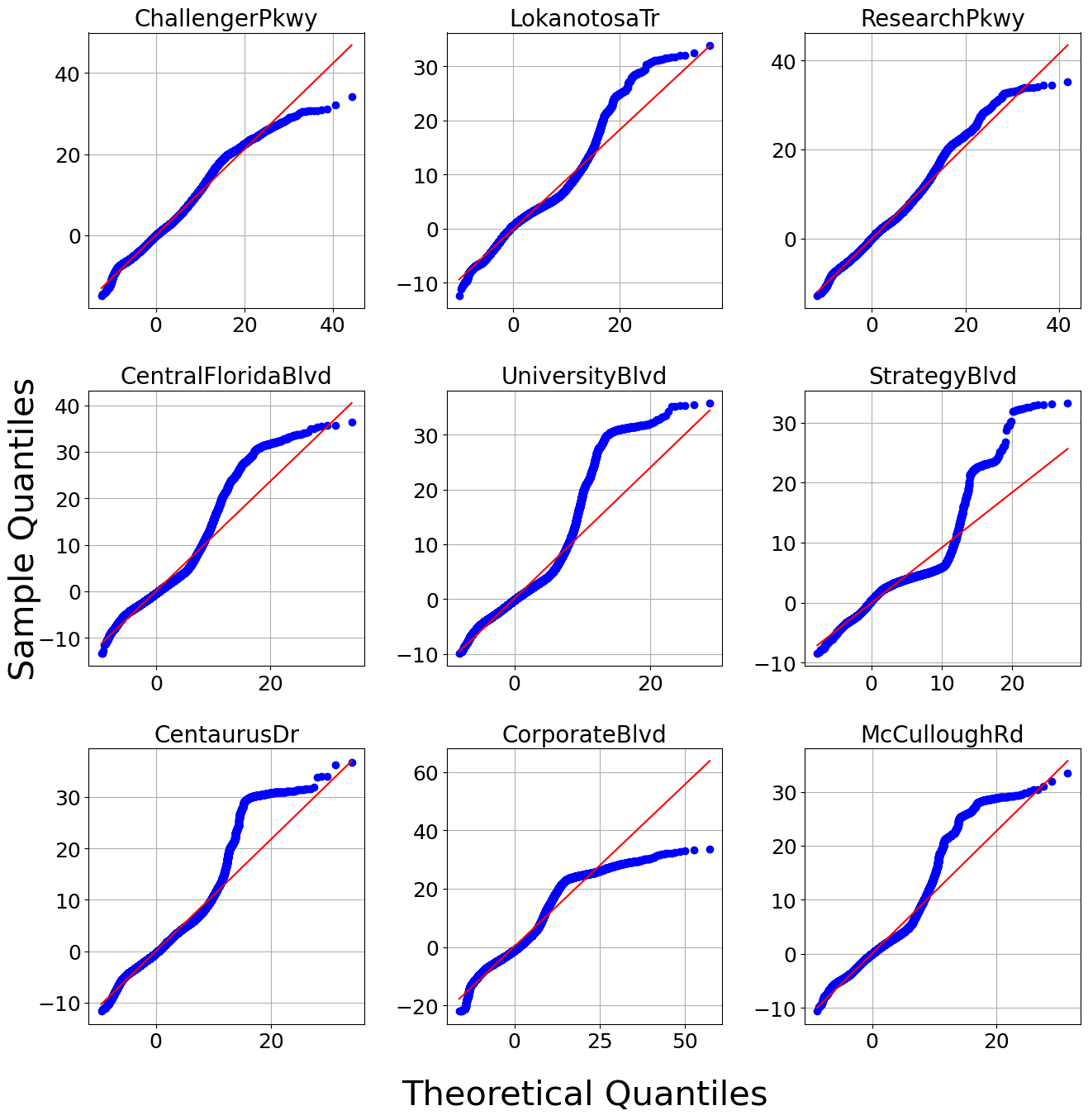}
        \caption{Q-Q plots of the data assuming it follows Gumbel distribution}
        \label{fig:qq}
    \end{figure}

% Python code links
% https://medium.com/the-researchers-guide/finding-the-best-distribution-that-fits-your-data-using-pythons-fitter-library-319a5a0972e9
\section{Conclusion}
The paper discusses the relationship between the cycle-to-cycle maximum queue length and extreme value theory. Characterizing long-term queue length patterns as GEV could lead to potential applications in arterial management. However, this paper only serves as a first step in utilizing EVT for queue length prediction and corridor management. In future studies, we plan to demonstrate the practical applications of this idea. Additionally, the paper only used the block maxima approach to characterize the extreme behavior of queue lengths. In future work, we plan to show the threshold exceedance analysis.

\bibliographystyle{IEEEtran}
\bibliography{EVT_ref}
\end{document}